Adrien de Jarmy
Doctorant Sorbonne Université, Orient & Méditerranée, UMR 8167

# Éléments de méthode : statistiques et humanités numériques. Quelles perspectives de recherche pour l'histoire des débuts de l'Islam ?

RESUME : Le développement des humanités numériques a ouvert de nouvelles perspectives en histoire de l'Islam : que l'on soit face à des corpus de sources minces ou parfois gigantesques (*Sīra*, al-Ṭabarī, al-Ḏahabī etc), ces outils nous permettent d'aborder bien plus efficacement les textes sous l'angle statistique, afin de conforter des hypothèses plus générales et dépasser les études de cas. En nous fondant sur les travaux d'un certain nombre de chercheurs et les nôtres, nous proposons d'étudier les potentialités et les limites posées par les méthodes informatiques pour l'histoire de l'Islam médiéval, en gardant à l'esprit que les conclusions pourront s'avérer utiles aux chercheurs d'autres périodes. Nous insisterons particulièrement sur deux outils : la construction et l'emploi des bases de données relationnelles, et plus récemment, le marquage ou *tagging* des sources, dont l'objectif affiché est de remédier à certains des problèmes posés par la précédente méthode.

MOTS-CLE : bases de données relationnelles ; tagging ; histoire de l'islam, islam médiéval ; histoire quantitative, Humanités numériques, méthodologie historique.

Lien vers l'article : http://www.revue-circe.uvsq.fr/elements-de-methode-statistiques-et-humanites-numeriques-quelles-perspectives-de-recherche-pour-lhistoire-des-debuts-de-lislam/

## Sommaire

Adrien de Jarmy
Doctorant Sorbonne Université, Orient & Méditerranée, UMR 8167

# Éléments de méthode : statistiques et humanités numériques. Quelles perspectives de recherche pour l'histoire des débuts de l'Islam ?

## Du problème de la valeur et de l'analyse des sources littéraires à grande échelle

Les difficultés que rencontre l'historien à la lecture des sources littéraires des débuts de l'Islam sont similaires aux questions suscitées par le caractère tardif des textes du Haut Moyen Âge européen. Si l'exemple des études néotestamentaires, par la voie de la méthode historico-critique revient régulièrement comme comparaison[1], on peut ainsi penser au problème auquel fait face l'historien qui étudie la figure de Clovis (m. 511) à travers la lecture de l'*Histoire des Francs* de Grégoire le Grand (m. 604), voire du *Liber historiae Francorum*, dont la plus ancienne copie date au plus tôt de 727. De même, pour reconstituer l'histoire du ier/viie[2] au Proche-Orient, telle la biographie historique du Prophète Muḥammad, les règnes des premiers califes dits *rāšidūn*[3] et le grand mouvement de conquêtes qui leur est associé, nous ne disposons d'aucun texte qui soit antérieur au début du iiie/ixe siècle, soit près d'un siècle et demi après les événements[4]. Cet important écart temporel qui sépare le temps des événements du temps de l'écriture des textes a engendré des débats houleux sur la valeur historique des sources littéraires des débuts de l'Islam. L'historiographie reste particulièrement divisée sur ce sujet, entre des travaux qui suivent toujours le récit de la Tradition islamique[5] dans ses grandes lignes et le révisionnisme de l'histoire de l'Islam, telle qu'elle a été canonisée par les oulémas à l'époque médiévale, en passant par quantités de positions intermédiaires. Ceci nous rappelle qu'il s'agit là avant tout d'un problème fondamental d'épistémologie, car qu'au-delà de l'interprétation que

---

[1] Voir par exemple *A Marginal Jew : Rethinking the Historical Jesus*, 5 vols., New-York, Yale University Press, 1991-2016, œuvre monumentale en cinq volume de John P. Meier.
[2] Les dates sont données selon le double calendrier hégirien/grégorien.
[3] Littéralement « bien-guidés » selon l'orthodoxie sunnite. Il s'agit d'Abū Bakr (m. 13/634), ʿUmar (m. 23/644), ʿUṯmān (m. 35/656) et ʿAlī (m. 40/661).
[4] L'exemple type est celui de la célèbre biographie d'Ibn Isḥāq (m. 150/762), rédigée déjà bien tardivement, et dont nous ne disposons plus que d'une transmission complète (*riwāya*) rédigée de la main d'Ibn Hišām (m. 213/828 ou 218/833), à partir d'une copie d'al-Bakkaʾī de Kūfa (m. 182/799).
[5] Ensemble des textes qui forment la *Sunna*, les règles de la loi islamique, dont la majeure partie a été canonisé autour des *ḥadīṯ*-s prophétiques. On y inclut aussi les textes plus narratifs, comme les récits des premières expéditions du Prophète Muḥammad, les *maġāzī*.

Adrien de Jarmy
Doctorant Sorbonne Université, Orient & Méditerranée, UMR 8167

l'on peut accorder aux événements, c'est bien notre capacité à accéder à la connaissance du passé qui est mise en jeu[6].

Le iiie/ixe siècle marque le début d'un fort accroissement de la production de textes, poussé par le développement des réseaux de savants dans les principales villes de l'Empire abbasside[7], l'utilisation du papier qui devient chose commune[8], ainsi que le processus de centralisation de l'État abbasside, dont la cour et les politiques de patronage attirent un plus grand nombre de lettrés[9]. C'est le temps de la rédaction des grandes vies du Prophète[10] et des recueils de ses faits et gestes classés par thèmes (les recueils de ḥadīṯ-s)[11], des histoires universelles[12] et des premières encyclopédies[13]. À partir de cette époque, la masse des sources pour étudier l'Empire abbasside contraste fortement avec le peu de données dont nous disposons pour les époques antérieures.

Pour étudier l'histoire des premiers siècles de l'Islam, mis à part les sources matérielles, l'historien est donc bien contraint de se pencher sur des sources littéraires produites à l'époque abbasside. Face à ce problème, les travaux académiques ont bien souvent été concentré sur des études de cas, sur un échantillon de textes dont on tirait des conclusions générales. Il est révélateur de noter que, concernant l'histoire des premiers siècles, les partisans d'une lecture

---

[6] Les grands moments de ce débat sont scandés par la remise en question progressive de la valeur historique des sources littéraires des débuts de l'Islam. Ce processus débute réellement à la fin du xixe siècle avec la publication des travaux d'Ignaz Goldziher, puis s'accélère entre les années 1950 et les années 1980, avec la publication d'une série d'études qui concernent d'abord le droit musulman (Joseph Schacht) puis l'ensemble des sources littéraires de la Tradition islamique (John Wansbrough, Patricia Crone, Martin Hinds). Les dernières théories révisionnistes surgissent au début des années 2000 avec la publication d'ouvrages tels que celui de Yehuda D. Nevo et Judith Koren, *The Crossroads to Islam*, New York, P. Books, 2003, dont l'étude numismatique reste cependant d'une grande utilité. Pour un éventail complet des positions et de la bibliographie, se reporter à Herbet Berg (dir.), *Method and Theory in the Study of Islamic Origins*, Leyde, Brill, 2003.

[7] The Princeton Encyclopedia of Islamic Political Thought, ed. Gerhard Bowering, Patricia Crone, Wadad Kadi, Devin H. Stewart, Muhammad Qassim Zaman, Mahan Mirza, Princeton, Princeton University Press, 2012, Joseph Schacht, *The Origins of Muhammadan Jurisprudence*, Oxford, Oxford University Press, 1959.

[8] Frederick Harrison, *A Book about Books*, Londres, John Murray, 1943. p. 79, Robin Myers et Michael Harris (ed). *A Millennium of the Book: Production, Design & Illustration in Manuscript & Print, 900–1900*, Winchester, St. Paul's Bibliographies, 1994. p. 182.

[9] Lawrence I. Conrad « The mawālī and aarly Arabic historiography" dans Monique Bernards et John Nawas, Patronate and Patronage in Early and Classical Islam, Brill, Leyde, 2005, p. 370-425.

[10] On pense à nouveau à la fameuse *Sīra* d'Ibn Isḥāq, dans sa recension par Ibn Hišām, mais aussi au *Kitāb al-maġāzī* ou Livre des expéditions d'al-Wāqidī (m. 207/823), connu par la recension de son secrétaire et disciplie Ibn Saʿd (230).

[11] Dès les premières décennies, ʿAbd al-Razzāq b. al-Ṣanʿānī (m. 211/827) et Ibn Abī Šayba (m. 234/849) rédigent deux grands recueils, les *muṣannaf*-s, dont les traditions prophétiques ne forment qu'une assez maigre partie. Les recueils canonisés par la tradition islamique, presqu'entièrement focalisés sur le Prophète et jugés les plus authentiques dans la Tradition islamiques sont les *kutub al-sitta*, la série de six recueils, dont al-Buḫārī (m. 256/870) est à la fois l'auteur le plus ancien et le plus éminent.

[12] La *Sīra* aurait pu inclure des traditions remontant à la naissance du monde dans sa première version. Mais on pense surtout au *Taʾrīḫ* d'al-Ṭabarī (m. 310/923), aussi connue en français sous le titre les Annales, dont le titre rappelle la comparaison historiographique avec Tacite (m. 120).

[13] Par exemple, les *Ṭabaqāt al-kubra* d'Ibn Saʿd, recueil de notices biographiques des grands hommes, débutant elle aussi par une compilation de traditions prophétiques.

Adrien de Jarmy
Doctorant Sorbonne Université, Orient & Méditerranée, UMR 8167

plus confiante envers la Tradition islamique, comme les plus critiques, se sont régulièrement accusés de procéder ainsi, comme cela a été le cas dans l'étude du ḥadīṯ par exemple[14]. Le gigantisme de certaines sources, tels les recueils de ḥadīṯ-s, les encyclopédies ou les histoires universelles, à l'image du Taʾrīḫ d'al-Ṭabarī, a pu aussi pousser au débat d'érudition, empêchant de porter un regard synthétique sur les textes et de détecter des grandes tendances.

Comment sortir du problème de l'étude de cas pour monter en généralité, et ainsi porter un jugement critique sur toute une œuvre ou même un corpus ? La méthode statistique dans les sciences historiques, dont le but est d'établir des données chiffrées sur les textes afin de donner un support positif et mathématique aux intuitions du chercheur, renoue avec l'esprit de l'école des Annales par le truchement du changement d'échelle. Elle rend plus sûr la détection des thèmes communs entre les textes et fournit un outil supplémentaire pour leur datation, à condition de donner sens aux données chiffrées en contextualisant l'apparition des thèmes dans les textes. Plus généralement, elle permet d'asseoir sur de solides fondations la compréhension du processus de canonisation des textes au sein d'un corpus, entre les récits qui peu à peu gagnent valeur d'autorité, et ceux qui sont laissés de côté. En cela, la méthode statistique établit aussi un pont avec la méthode historico-critique allemande, et notamment la méthode de la critique de la rédaction[15].

Ce travail est aujourd'hui grandement facilité par l'emploi des outils numériques. Concernant l'histoire de l'Islam médiéval, de nombreuses bases de données qui prennent la forme de recueils de textes sont désormais accessibles et permettent d'effectuer des recherches rapides ainsi que des estimations statistiques manuelles relativement simples[16]. Ne nécessitant que des compétences informatiques basiques, cet outil est le plus largement employé

---

[14] Voir par exemple les critiques adressées par Muhammad Qassim Zaman dans *Religion and Politics under the early ʿAbbāsids. The Emergence of the Proto-Sunnī Elite*, Leyde, New-York, Cologne, Brill, 1997 envers l'ouvrage de Patricia Crone et Martin Hinds dans *God's Caliph: Religious Authority in the First Centuries of Islam*, London, Cambridge University Press, 1986, leur reprochant de fonder leur analyse de la titulature des califes omeyyades sur des extraits de la poésie de cours d'al-Farazdāq (m. 11/730). Inversement Stephen J. Shoemaker, partisan d'une approche très critique des sources dans *The Death of a Prophet: The End of Muhammad's Life and the Beginning of Islam*, Philadelphie, The University of Pennsylvania Press, 2011, a mis en doute la méthode « isnād-cum-matn » développée par Harald Motzki dans « Dating Muslim Traditions: A Survey », *Arabica*, 2005, 2/52, p. 204-253, dont le but est de dater l'apparition des traditions prophétiques dans les textes en procédant à la comparaison de plusieurs transmissions pour chacune d'entre elle, en comparant à chaque fois les deux éléments qui les caractérisent : la chaîne des garants (*isnād*) et le récit de la tradition, c'est-à-dire le contenu du texte lui-même (*matn*).

[15] *Redaktionsgeschichte*. Méthode d'exégèse et de datation des textes bibliques dont une des composantes principales repose sur l'analyse des répétitions thématiques dans un même texte.

[16] Quelques-unes, lesp lus généralistes, sont bien connues des arabisants. C'est le cas d'*al-Maktaba al-Shamela* (shamela.ws), d'*al-Warraq* (www.alwarraq.net) ou d'*al-Meshkat* (www.almeshkat.net). Il en existe pourtant bien d'autres, selon le type de corpus, telle la *Noor Digital Library* pour les textes chiites (www.noorlib.ir), ArabiCorpus (arabicorpus.byu.edu) qui propose des outils lexicographiques, à titre d'exemples.

Adrien de Jarmy
Doctorant Sorbonne Université, Orient & Méditerranée, UMR 8167

aujourd'hui. Les bases de données relationnelles, dont l'objectif est de relier entre elles des données préalablement insérées par l'utilisateur, sont bien plus rarement utilisées. L'objectif n'est pas le même puisqu'il s'agit ici, non pas de rechercher manuellement des références dans un texte, mais d'automatiser les estimations statistiques en lançant des requêtes informatiques dans la base[17]. La première phase du travail est très laborieuse, puisque l'utilisateur doit insérer les données qui seront ensuite destinées à être traitées, et il doit pour cela décider de s'investir personnellement dans l'apprentissage de l'outil. En revanche, il peut personnaliser très précisément le type de recherche qu'il veut lancer dans la base de données, en établissant des critères de tri ou de croisement des sources, et établir des estimations statistiques bien plus complexes et plus sûres. C'est pourquoi, en nous fondant sur une méthode développée pour notre travail de thèse et que nous développons dans la partie suivante[18], nous proposons dans cet article d'étudier les potentialités offertes par les bases de données relationnelles pour l'établissement de statistiques et l'analyse des grandes thématiques des textes dans l'histoire des débuts de l'Islam. Nous montrerons qu'il est alors possible d'automatiser la recherche de thèmes ou d'auteurs communs entre les textes et d'évaluer statistiquement les relations entre différentes sources. En pointant aussi les apories de ce travail, on s'intéressera aux outils numériques qui visent à combler les limites techniques des bases de données relationnelles, notamment à-travers la méthode du *tagging*. En développant une réflexion générale sur ces outils et leur potentiel statistique, nous espérons pouvoir établir un pont avec les chercheurs travaillant sur d'autres aires géographiques et temporelles, voire d'autres disciplines.

# Réflexion épistémologique sur les modalités de construction d'une base de données relationnelle

La construction d'une base de données relationnelle suppose de réfléchir en amont aux catégories dans lesquelles les données sont insérées, et donc à la nature des informations et des sources exploitées par le chercheur. De ce questionnement proprement technique émerge une réflexion épistémologique, puisque celui-ci doit anticiper les données qui sont susceptibles de ressortir de la base. Il s'interroge ainsi aux moyens d'accès à la connaissance historique,

---

[17] Voir Annexe 1 pour une illustration d'une base de données relationnelle. Toutes les tables sont reliées entre elles afin de lancer des recherches croisées dans la base.
[18] Méthode que nous développons dans notre thèse : « La construction historiographique de la figure du Prophète dans les sources des débuts de l'Islam, ier/viie-ive/xe siècle ».

Adrien de Jarmy
Doctorant Sorbonne Université, Orient & Méditerranée, UMR 8167

l'encourageant à repenser des catégories parfois considérées comme relevant de l'évidence dans sa discipline[19].

Le chercheur doit au préalable se familiariser avec un langage de programmation. Bien que de nombreux logiciels proposent un assistant à la création de la base de données relationnelle afin d'éviter à avoir à la programmer manuellement – c'est le cas par exemple de Base, logiciel de la suite OpenOffice[20] –, la connaissance d'un langage évite bien des surprises lors de l'enregistrement des données par la suite[21]. Le SQL, abréviation de *Structured Query Language*, est aujourd'hui le langage standard d'accès aux bases de données relationnelles.

La première étape consiste ensuite dans le choix des tables et des catégories, là débute réellement la réflexion épistémologique. La table correspond au premier niveau de l'architecture de la base de données. Elle permet déjà, dans un premier temps, de classifier les informations en grands types. Dans le cadre de notre thèse, nous avons ainsi créé un certain nombre de tables qui répondent directement à la structure des sources de la Tradition islamique. En effet, tout corpus de traditions dans les premiers siècles de l'Islam s'organise selon un schéma similaire : les recueils de textes – qu'il soit question des recueils de *ḥadīṯ*-s ou des œuvres suivant un fil chronologique, tel la *Sīra* du Prophète – sont divisés en nombreuses unités narratives. Chacune d'elle comporte une chaîne de garants, c'est-à-dire l'identité des individus censés avoir transmis le récit et dans le but de prouver son authenticité (*isnād*)[22], et le récit, c'est-à-dire l'information (*matn*)[23]. La Tradition islamique, malgré des données essentiellement qualitatives et narratives, est marquée par une structure itérative et se prête donc bien à la recension dans les catégories des bases de données. La subdivision du texte entre l'*isnād* et le *matn* dans la très grande majorité des textes facilite le processus. On peut extrapoler cet exemple

---

[19] Dans notre cas, alors que les études sur le *ḥadīṯ* et les *maġāzī* relèvent traditionnellement de champs disciplinaires distincts, nous avons décidé d'effectuer des requêtes croisées entre ces sources, en nous fondant sur des arguments formels. Les requêtes font alors apparaître des thématiques communes aux textes.
[20] Les annexes de cet article sont tirées du logiciel.
[21] Plusieurs cours sont disponibles afin d'apprendre le langage SQL. Nous nous sommes aidés des formations proposées sur le site https://openclassrooms.com. Pour les historiens, le site https://programminghistorian.org/en/lessons/ propose également un certain nombre de cours (en anglais) et plusieurs outils de qualités pour extraire les données de la base. Il est également possible de se référer aux nombreux outils proposés dans *Historical Research Handbook: Designing Databases for Historical Research* (https://port.sas.ac.uk/mod/book/view.php?id=75)
[22] Chaîne de noms sensée remonter au premier locuteur de l'information.
[23] Exemple de la structure d'une tradition. *isnād* : al-Wāqidī ← Abū Bakr b. Ismāʿīl ← Ismāʿīl ← ʿĀmir b. Saʿd ← Saʿd. Résumé du matn ou du texte de la tradition : ʿUmayr b. Abī Waqqāṣ, adolescent au sein de l'armée du prophète, se cache dans les rangs de ses aînés, de peur que celui-ci ne lui refuse d'aller au combat à cause de son jeune âge, et le prive ainsi d'accéder au Paradis en qualité de martyre. al-Wāqidī, *Kitab al-maġāzī*, ed. Marsden Jones, Londres, Oxford University Press, p. 22.

Adrien de Jarmy
Doctorant Sorbonne Université, Orient & Méditerranée, UMR 8167

à d'autres types de sources, l'important étant de concevoir la table à partir d'un ensemble de critères communs entre les textes.

À titre d'exemple, nous avons ainsi programmé une table permettant de recenser l'identité des individus[24], une table servant à enregistrer l'ensemble du contenu de la tradition[25] ainsi qu'une table dressant la liste des sources étudiées[26]. Toutes ces tables sont reliées entre elles par des tables de relations, dont le but n'est autre que de lier les tables primaires les unes entre les autres afin de permettre à l'utilisateur de réaliser des requêtes croisées dans l'ensemble de la base de données[27]. Toute information enregistrée dans l'une des deux tables est bien entendue chiffrée pour être compréhensible par l'ordinateur[28], ce nombre sert à identifier l'information. Toute nouvelle liaison entre les deux tables produit aussi un nouveau numéro servant à identifier la connexion entre les données des deux tables[29].

La seconde étape concerne le choix des catégories inhérentes aux tables, c'est en quelque sorte, une sous division des tables. Trois grands types de catégories sont possibles :

- Les catégories booléennes[30]. Ce sont les catégories où l'utilisateur n'a le choix qu'entre trois types de réponses : oui (vrai), non (faux) ou la superposition des deux, la troisième réponse correspondant en logique à cet état de superposition qu'est le chat de l'expérience de Schrödinger, évitant de remettre en cause le principe du tiers exclu[31]. Dans notre étude, nous avons ainsi sélectionné des catégories très générales permettant de classer les traditions en grands thèmes : cette tradition a-t-elle une thématique militaire ? théologique ? fiscale ? Dans ce type de cas, seules les réponses oui ou non sont envisageables[32]. Si l'on prend l'exemple de la tradition précédemment citée pour illustrer la structure d'une tradition islamique[33], alors on comprend qu'il s'agit d'un texte ayant une portée militaire, et il suffira de cocher la case appropriée pour que celle-ci ressorte dans une future requête qui viserait à lister toutes les traditions rangées sous cette catégorie dans la base de données. Il est bien entendu possible de cocher plusieurs

---

[24] Annexe 1, TBL–indiv.
[25] Annexe 1, TBL–trad.
[26] Annexe 1, TBL–recueil.
[27] Annexe 1, TBL–indiv–recueil, TBL–indiv–trad et TBL–recueil–trad.
[28] Annexe 2, voir les données chiffrées des colonnes id–indiv et id–trad.
[29] Annexe 2, id–indiv/trad.
[30] Du nom du mathématicien et logicien britannique George Boole (m. 1864). Annexe 3. Les catégories booléennes sont des cases à cocher afin de faire entrer la tradition dans une catégorie à choix fermé.
[31] Erwin Schrödinger, *Physique quantique et représentation du monde*, Paris, Le Seuil, 1992, p. 184.
[32] Voir à nouveau annexe 3.
[33] Voir note 27.

Adrien de Jarmy
Doctorant Sorbonne Université, Orient & Méditerranée, UMR 8167

cases si une tradition appartient à plusieurs catégories. Pour d'autres catégories plus restreintes, la troisième solution peut être d'une grande utilité. Prenons cet exemple : cette tradition comprend-elle un miracle ? Le Prophète fait surgir un torrent d'eau en plein désert pour venir en aide à ses Compagnons, comme c'est le cas dans le récit de l'expédition de Tabūk dans plusieurs sources, il suffit alors de cocher la première case pour retrouver par la suite lors d'une requête, toutes les traditions qui mentionnent un miracle du Prophète. La tradition n'en fait aucunement mention, la seconde option est alors évidente. Mais il peut arriver que la situation soit moins claire. Que choisir lorsque l'événement est sous-entendu mais non explicité ouvertement ? Lorsqu'une tradition plus archaïque nous donne les premiers éléments surnaturels du récit, sans pour autant expliciter clairement qu'il s'agit d'un miracle, que le mot n'est pas prononcé ? Le récit est à ce moment dans un entre-deux, reflétant l'état intermédiaire d'un processus qui associe cet événement à un miracle de la vie du Prophète. La troisième solution permet alors de penser cet entre-deux et de le localiser par la suite dans la base de données.

- Choix multiple. Ce type de catégorie permet de sélectionner un choix parmi un ensemble de réponses préétablies dans la base de données. Nous avons fait le choix de cette catégorie pour programmer une liste de tous les noms qui apparaissent dans la chaîne des garants des traditions. Ceci nous permet ensuite d'établir une requête pour retrouver l'ensemble des traditions transmises sous un nom particulier[34], ou même de croiser des listes de noms afin de reconstruire les suites les plus courantes. De manière générale, les catégories à choix multiple facilitent grandement l'établissement de listes de tout type ainsi que leur étude au sein d'un corpus.

- VARCHAR ou TEXT. Ce type de catégorie ne permet pas de réaliser de requête automatique, il s'agit simplement ici d'enregistrer du texte de taille variable selon les besoins. Nous nous en servons de notre côté pour noter le résumé des traditions et éviter de devoir retourner systématiquement au texte de la source.

On comprend ainsi que chaque type de catégorie remplit un but bien précis, qu'il s'agisse de classer une tradition parmi un nombre défini de catégories préétablies ou de lui attribuer une identité en incluant la liste des transmetteurs de l'information. Tout l'intérêt de la base de

---

[34] Annexe 4.

Adrien de Jarmy
Doctorant Sorbonne Université, Orient & Méditerranée, UMR 8167

données relationnelle réside dans la possibilité de croiser les catégories entre elles en lançant une recherche, et d'estimer mathématiquement ces liaisons en prenant appui sur des informations statistiques.

# Potentialités d'une base de données relationnelle en matière de datation et de processus de canonisation au sein des sources textuelles

Dans le cadre de notre travail de thèse, l'objectif est d'obtenir des statistiques sur le pourcentage de traditions qui mentionnent le Prophète Muḥammad dans notre corpus, afin d'évaluer l'évolution de son autorité dans les textes, et dater l'émergence de cette figure dans la littérature des premiers siècles de l'Islam. S'il est possible d'obtenir ces données manuellement, la base de données relationnelle permet d'automatiser la recherche et de croiser les résultats avec d'éventuelles futures requêtes. En suivant les différentes étapes citées précédemment, nous avons pu produire un certain nombre de statistiques de notre base de données en nous fondant sur un échantillon de livres des expéditions[35] du Prophète extrait de recueils de ḥadīṯ-s[36]. Ici, l'intérêt d'obtenir des statistiques est d'étudier la progression d'un type de donnée enregistré dans la base de données au sein d'un corpus de textes. Dans le tableau de l'annexe 3, nous avons sélectionné trois textes que l'on peut dater de trois époques différentes[37] :

- Le livre des expéditions de ʿAbd al-Razzāq b. al-Ṣanʿānī (m. 211/827), inclus dans son *Muṣannāf*[38]. En réalité, l'entièreté du texte nous provient de son maître Maʿmar b. Rāšid (m. 152/770), historiographe à la cour des Omeyyades de Damas et qui s'installa à Ṣanʿā au Yémen après la prise de pouvoir des Abbassides en 132/750.
- Le livre des expéditions de l'irakien Ibn Abī Šayba (m. 234/849), lui aussi inclus dans son propre *Muṣannaf*.
- Le livre des expéditions d'al-Buḫārī (m. 256/870), inclus dans son *Ṣaḥīḥ*, premier recueil de tradition de la série des recueils canoniques de la Tradition islamique[39].

---

[35] *Kutub al-maġāzī*, littéralement « livre » ou tout document écrit recensant « les lieux des expéditions » du Prophète Muḥammad.

[36] Voir annexe 5. Ce tableau est extrait d'un article à paraître dans le premier volume du projet "The Presence of the Prophet: Muḥammad in the Mirror of His Community in Early Modern and Modern islam": Adrien de Jarmy, « Dating the Emergence of a Warrior-Prophet Character in the *Maġāzī* Literature (2nd/8th – 4th/10th century) », dans Denis Gril et Stefan Reichmuth (dir.), "Representations of the Prophet in Doctrine, Literature and Arts", Handbuch der orientalistik, Leyde, Brill, 2021.

[37] En l'absence d'indication sur la date de rédaction du texte, la convention est de se référer à la date de mort du traditionniste comme date *terminus post quem*.

[38] Compilations de traditions classées par thèmes, comme l'indique la racine arabe *ṣa-na-fa*.

[39] Sur le processus de canonisation de cette œuvre majeure par les oulémas de l'Empire abbasside, voir Jonathan Brown, *The Canonization of al-Bukhārī and Muslim: the Formation and Function of the Sunnī ḥadīth Canon*, Leyde, Brill, 2007.

Adrien de Jarmy
Doctorant Sorbonne Université, Orient & Méditerranée, UMR 8167

En réalisant une recherche dans la base de données, nous avons pu établir trois niveaux d'analyse :
- Le nombre de traditions dans chaque source.
- Le nombre absolu et le pourcentage de traditions mentionnant le nom du Prophète dans chacun des textes.
- Le nombre absolu et le pourcentage de chapitres qui mentionnent son nom.

Une simple requête dans la base de données permet à l'utilisateur d'avoir une idée assez précise de la répartition du type de donnée sélectionné dans le corpus. Premièrement, nous pouvons tout de suite voir l'évolution générale du nombre de traditions recensées dans chacun des textes. Alors que le texte de Maʿmar ne comprend que 147 traditions au total, les deux traditionnistes suivants en compilent respectivement 579 et 488. On comprend donc qu'à l'époque du premier traditionniste, les textes de *maġāzī* sont encore à un stade de développement précoce. Le texte est bien plus court que les sources de ce type que l'historien rencontre habituellement à l'époque abbasside. Au contraire, les deux textes d'Ibn Abī Šayba et d'al-Buḫārī indiquent une nette augmentation dans les deux premiers tiers du iiie/ixe siècle, suggérant un processus de maturation des textes. De simples petites compilations de traditions au tournant de la Révolution abbasside, les livres des expéditions deviennent d'importants recueils de tradition au siècle suivant. Bien que cela soit déjà mentionné par les spécialistes de la question[40], la méthode statistique permet de confirmer une intuition en faisant reposer l'argumentation sur des preuves plus solides. Mais il s'agit aussi d'évaluer la progressivité de cette construction en définissant clairement les étapes de cet effort de compilation et en reliant ces données au contexte de production de la source et à la vie de son auteur. Maʿmar b. Rāšid, ancien lettré à la cour des Omeyyades, trouve refuge à Ṣanʿā au Yémen, après la conquête abbasside en 132/750 et transmet les traditions compilées à son disciple favori, ʿAbd al-Razzāq b. al-Ṣanʿānī. La base de données relationnelle permet d'envisager avec plus de précision l'état de l'historiographie à la fin de la période omeyyade, alors même que nous ne disposons pas du texte.

---

[40] Pour un résumé de la question voir Martin Hinds, « al-Maghāzī », *EI²* et *«Maghāzī» and «Sīra» in early Islamic scholarship*, dans *La vie du Prophète Mahomet*, colloque de Strasbourg, 23-24 octobre 1980, Paris 1983, 57-66, ainsi que les deux études sur le sujet : Josef Horovitz, *The earliest biographies of the Prophet and their authors*, éd. Marmaduke Pithall, *Islamic Culture*, 1927, p 537-539 et Rudi Paret, *Die legendäre Maghāzī-Literatur: arabische Dichtungen über die muslimischen Kriegzüge zu Mohammeds Zeit*, Mohr, Tübingen, 1930.

Adrien de Jarmy
Doctorant Sorbonne Université, Orient & Méditerranée, UMR 8167

Deuxièmement, on peut se pencher sur les pourcentages extraits de la base de données. Ces pourcentages nous renseignent sur l'évolution générale des traditions prophétiques, et donc sur la place de la figure du Prophète Muḥammad dans les textes. Une lente évolution se dégage ainsi de l'analyse comparative des sources : de 61,9% de traditions prophétiques dans le premier texte, on passe à 88,3% un siècle plus tard, dans le Ṣaḥīḥ d'al-Buḫārī. On comprend aussi que cette évolution n'est pas linéaire, puisqu'à l'époque d'Ibn Abī Šayba, dans la première moitié du iiie/ixe siècle, le pourcentage de traditions prophétiques n'atteint plus que 55,1% du texte, un peu plus seulement de la moitié. La requête nous permet également de mesurer la diffusion des traditions prophétiques à l'échelle de chacun des chapitres. Celle-ci est bien plus constante, de 73% dans le premier texte, on passe à 100% dans le second et 96,6% dans le dernier. Chaque chapitre correspond à une sous-thématique ou à un événement important dans la chronologie de l'histoire des débuts de l'islam. Autrement dit, la diffusion des traditions prophétiques est déjà entièrement actée dans les premières décennies du iiie/ixe siècle et colonise l'ensemble des sujets abordés dans ces textes. On comprend que l'autorité du Prophète, aujourd'hui un unanime parmi tous les courants de l'islam, est pourtant loin d'être évidente et reflète une construction progressive. À la fin de l'époque omeyyade, cette autorité est encore en construction et ne fait pas l'objet d'un consensus, tout une partie du texte ainsi qu'un nombre important de chapitres ne font aucune mention du Prophète. La lecture des catégories VARCHAR/TEXT de la base de données nous révèle qu'il est en fait bien souvent question d'événements postérieurs à la Révélation prophétique, notamment concernant les premiers califes dits bien-guidés. Sur la base de ces chiffres et d'une lecture attentive des textes, on peut formuler l'hypothèse que les débuts de l'historiographie islamique sont marqués par l'émergence de plusieurs autorités, le Prophète ne prenant le dessus que très progressivement sur les autres. Les raisons qui amènent les traditionnistes à favoriser cette figure plutôt qu'une autre feront l'objet d'une discussion dont nous traiterons dans le corps de notre travail de thèse.

La méthode statistique, épaulée par les outils numériques, fournit également de solides fondations pour une datation de ces représentations. Si l'on peut bien entendu toujours discuter des chiffres qui n'ont aucune valeur en eux-mêmes, cette méthode permet d'éviter un écueil important qui touche malheureusement régulièrement ce que l'on nomme parfois l'histoire des représentations et les problèmes méthodologiques qui y sont associés : la multiplication en série des représentations, décontextualisées de leur milieu et de toute forme de datation. Alors que notre travail de thèse a pour but de dater les représentations du Prophète dans les premiers textes de l'islam médiéval, on peut imaginer bien d'autres utilités pour cette méthode, selon le type de donnée que l'on souhaite faire apparaître dans les requêtes. À ce titre, une seconde requête,

Adrien de Jarmy
Doctorant Sorbonne Université, Orient & Méditerranée, UMR 8167

fondée sur la liste de noms (*isnād*-s), nous as permis de relier l'ensemble des traditions – à l'exception d'une seule – présentes dans le livre de ʿAbd al-Razzāq au nom de son maître Maʿmar, facilitant leur datation et l'automatisation des recherches[41]. Ce processus sera répété dans toutes les sources de notre corpus afin de reconstruire la liste des *isnād*-s, c'est-à-dire la généalogie des autorités qui revendiquées la transmission de traditions prophétiques dans la Tradition islamique des premiers siècles.

La réalisation d'une base de données relationnelle facilite ainsi le croisement de tout un ensemble de catégories et de thématiques pour automatiser la recherche, rejoignant une nouvelle approche historico-critique des sources. Les potentialités sont particulièrement intéressantes lorsque les textes partagent une structure similaire, la base de données prend ainsi la forme d'une grande liste, suivant la forme d'une pyramide[42], dont les nombreuses catégories et sous-catégories forment les blocs. On peut cependant déjà souligner un problème majeur inhérent à la méthode : sa relative rigidité. En effet, la construction d'une base de données relationnelle suppose au préalable d'avoir une idée relativement précise des sources, afin de modeler sa structure en fonction du type de données que l'on souhaite enregistrer. La base de données ne permet de chercher que des types de données préalablement déjà connues du chercheur. Notre étude a été largement facilité par une série de sondage effectués lors de nos travaux de master, cependant, la multiplication des inattendus aurait pour corollaire la multiplication des catégories, rendant particulièrement complexe la requête la plus simple. La méthode du *tagging*, dont le processus d'automatisation du marquage des textes arabes est en cours de développement, a pour objectif de pallier cette rigidité[43].

# La méthode du *tagging* comme réponse aux apories des bases de données relationnelles

La méthode quantitative reste malgré tout laborieuse. Comme le rappelle Maxim Romanov, le passage aux bases de données numériques a permis une certaine progression dans le tri des données, donnant aussi la possibilité à l'utilisateur d'automatiser les requêtes et comme nous

---

[41] Voir l'introduction de l'édition du texte par Sean W. Anthony, *The Expeditions. An Early Biography of Muḥammad by Maʿmar Ibn Rāshid according to the recension of ʿAbd al-Razzāq al-Sanʿānī*, New York, London, New York University press, 2014.

[42] Voir à nouveau la structure de la base de données relationnelle en annexe 1 ainsi que la liste en annexe 4.

[43] Maxim Romanov, Computational Reading of Arabic Biographical Collections with Special Reference to Preaching in the Sunni World (661-1300 CE), University of Michigan, 2013, p. 58-59.

Adrien de Jarmy
Doctorant Sorbonne Université, Orient & Méditerranée, UMR 8167

l'avons souligné précédemment, de croiser facilement les catégories. Cependant, sa construction reste il est vrai, très fastidieuse, l'enregistrement des données se fait toujours manuellement, ce qui n'exempte pas le chercheur de commettre occasionnellement quelques erreurs[44]. La méthode dite du *tagging*, c'est-à-dire du « marquage » en français, correspond au marquage numérique des termes au sein d'un corpus océrisé[45]. Il s'agit de *tagger*, c'est-à-dire marquer les termes que l'on souhaite enregistrer au sein d'une base de données – un fichier source – dont le fonctionnement est bien différent. L'incrémentation des données est ainsi réalisée au fur et à mesure de la lecture des sources, ou alors en procédant simplement à des recherches lexicales ciblées. Cette nouvelle base de données fonctionne alors comme un réseau neuronal de points, elle se construit de manière bien plus évolutive et permet de remédier aux deux grands problèmes des bases de données relationnelles : la chronophagie et la rigidité des catégories.

En effet, la construction d'une base de données relationnelles suppose que l'utilisateur, en réfléchissant aux catégories qu'il souhaite programmer, réfléchisse déjà à l'architecture des connaissances et à leur sens. La construction est donc déjà interprétation et présuppose un jugement. Elle forme un nombre restreint de possibilités dont la requête est la résultante. Au contraire, la base de données construite par *tagging* n'a aucun sens en elle-même, l'interprétation et le jugement ne prennent forme qu'au moment où la requête est lancée et où les relations entre les différents termes du lexique sont créées. L'utilisateur ne fait rien d'autre que de créer un lexique malléable à merci au sein de son corpus, ce qui est à la fois son plus grand atout mais aussi sa faiblesse : les mots du lexique sont extraits du contexte textuel et risquent de perdre leur sens. Il est ensuite possible de formuler toutes sortes d'associations entre des termes pour formuler bien entendu des statistiques à très grande échelle, mais également pour repérer des répétitions lexicales – que ce soit à l'intérieur d'un chapitre, d'une source ou de tout un corpus – ou pour définir le degré d'association entre les termes choisis[46]. Au-delà de l'analyse du vocabulaire, il est possible de repérer des constructions complexes, soit des

---

[44] Maxim Romanov, "Toward the Digital History of the Pre-Modern World: Developing Text-Mining Techniques for the Study of Arabic Biographical Collections", dans *Methods and Means for the Digital analysis of Ancient and Medieval Texts and Manuscripts. Proceedings of the Conference*, Louvain, Brepols, 2012. Voir cet article pour une description technique de la méthode, étape par étape.

[45] Un fichier texte dont chacun des caractères est reconnu en standard Unicode et dans lequel l'utilisateur peut effectuer des recherches lexicales. Pour une illustration, voir Maxim Romanov, Computational Reading of Arabic Biographical Collections with Special Reference to Preaching in the Sunni World (661-1300 CE), University of Michigan, 2013, p. 66.

[46] Pour avoir une idée des réalisations graphiques rendues possibles par cette méthode, voir la thèse de Maxim Romanov, *Computational Reading of Arabic Biographical Collections with Special Reference to Preaching in the Sunni World* (661-1300 C.E.), University of Michigan, 2013.

Adrien de Jarmy
Doctorant Sorbonne Université, Orient & Méditerranée, UMR 8167

expressions régulières en termes informatiques[47]. On peut imaginer qu'un tel procédé permette de repérer automatiquement toutes sortes de formulations dans les sources de l'islam médiéval : formulations coraniques, répétitions de *topoï* caractérisant les récits prophétiques ou autres – on retrouve là aussi la problématique de la canonisation des textes –, syntaxes caractéristiques des prières et admonestations[48] etc. Muazzam Ahmed Siddiqi et Mostafa l-Sayed Saleh ont déjà montré les possibilités d'une telle approche pour l'extraction automatique des chaînes de transmissions des textes, des *isnād*-s [49], ce qui permet auw auteurs de l'article de retrouver toutes les occurrences d'un nom en ayant préalablement « taggé » le mot dans le lexique de la base de données. De manière générale, le tagging permet de préparer un corpus particulièrement flexible pour la recherche. La méthode est bien sûr applicable à bien d'autres corpus de textes, tant que ceux-ci sont océrisés. En outre, elle ouvre la porte au potentiel offert par l'intelligence artificielle et donc à la possibilité que l'ordinateur sache repérer au fur et à mesure, toutes les constructions annexes aux requêtes demandées par l'utilisateur[50].

Bien entendu, la méthode du *tagging* reste un outil, elle ne fournit elle aussi que des données dont l'interprétation selon des critères académiques (historique, linguistique ou autre) dépend de la corrélation avec des données externes au corpus. Autrement dit, elle ne remplace par l'analyse et le jugement du chercheur. En effet, à l'heure actuelle, l'ordinateur qui reçoit des requêtes au sein d'une base de données construite par *tagging* n'est pas encore capable de différencier le contexte énonciatif dans lequel prennent place ces termes. Prenons un exemple fondé sur la structure des sources étudiées dans notre travail de thèse. Nous avons vu précédemment que les sources des débuts de l'islam sont marquées par la division entre l'*isnād*, la chaîne des garants, et le *matn*, c'est-à-dire le récit ou la narration. En suivant cette méthode,

---

[47] Le terme est dû au mathématicien et logicien Stephen Cole Kleene (m. 1994). Issue des théories mathématiques des langages formels, une expression régulière désigne une chaîne de caractères suivant une syntaxe précise un ensemble de chaînes de caractères possibles. En informatique, une expression régulière est en fait un code reconnaissable par l'ordinateur, la chaîne de caractères d'une langue donnée (caractères latin, arabes ou autres) correspondant à sa transcription en langage Unicode. Voir plus bas les travaux qui sont effectués sur des ensembles de mots pour former des expressions ou *Natural Language Processing* (NLP).
[48] C'est par exemple une piste évoquée par exemple par Guillaume Dye pour étudier le processus de canonisation du texte coranique dans « Le corpus coranique, questions autour de sa canonisation », dans *Le Coran des historiens*, Paris, Le Cerf, 2019, p. 859-906.
[49] Pour une illustration voir Muazzam Ahmed Siddiqi et Mostafa l-Sayed Saleh, « Extraction and Visualization of the Chain of Narrators from Hadiths using Named Entity Recognition and Classification », *International Journal of Computational Linguistisc Research*, 5/1, 2014, p. 14-25.
[50] Voir le projet ERC « Kitāb » et les données produites à partir des recherches du corpus arabe OpenITI : http://kitab-project.org, ainsi que le projet OpenITI mARKdown : https://alraqmiyyat.github.io/mARKdown/. Pour une introduction à ces outils pour la langue arabe voir Nizar Y. Habash, *Introduction to Arabic natural language processing*, New-York, New York University Press ,2010. Pour une utilisation du *tagging* dans d'autres contextes comme le grec ancient voir Giueseppe G. A. Celano, Gregory Crane, Saeed Majidi, « Part of Speech Tagging for Ancient Greek », *Open Linguistics*, 2/1, 2016 ainsi que les nombreux outils pour historiens développés par Michael Piotrowski sur son site NLP for Historical Texts, https://nlphist.hypotheses.org/author/mxpi.

Adrien de Jarmy

Doctorant Sorbonne Université, Orient & Méditerranée, UMR 8167

on pourrait être tenté de lancer une requête sous la forme d'une expression régulière, afin de repérer les listes de noms des *isnād*-s, et ainsi séparer automatiquement l'*isnād* du *matn*. Cependant, l'ordinateur ne fera pas la différence entre les noms insérés dans la liste des garants et ceux qui font partie du corps de la narration. L'utilisateur devra vérifier personnellement chacune des références repérées au sein du corpus, ce qui prendrait bien plus de temps que de réaliser une base de données relationnelle. Bien entendu, l'ordinateur n'est pas non plus capable de différencier deux personnes portant le même nom au sein d'un même texte. Or, les textes de l'islam médiéval sont remplis de personnages très sobrement nommés par leur prénom ou une suite similaire de surnoms. Appliquer sans distinction une requête de ce type induirait le chercheur à faire de grossières erreurs. Le second problème est plus spécifiquement lié aux éditions des textes médiévaux dans le monde musulman. Les éditions des sources canoniques, tels les grands recueils de *ḥadīṯ*-s par exemple, sont réellement foisonnantes. Les éditions océrisées, de certaines bases de données en ligne telle al-Maktaba al-Šāmla[51], sont de qualités très variables et ne sont pas exempts non plus de volonté de reformulation ou de tri selon un possible biais idéologique. Pour les éditions plus scientifiques, comme le corpus OpenITI, ce problème n'est pas non plus complètement résolu, puisqu'il peut exister des différences notables entre les différentes éditions d'un même texte. Pour y remédier, il faudrait ainsi lancer des requêtes sur l'ensemble des éditions d'un corpus de textes, ce qui a pour corollaire la démultiplication du travail d'océrisation.

## Conclusion

Qu'il s'agisse de construire une base de données relationnelle ou un fichier source à partir de la méthode du *tagging*, ce type de travail est idéalement à réaliser en équipe. En effet, le caractère laborieux du travail, mais aussi en retour, les potentialités offertes par les requêtes, conviennent particulièrement au partage du fardeau, comme de la récompense. Dans l'attente de l'évolution technique de la méthode du marquage, notamment grâce à la progression de l'intelligence artificielle, on peut envisager que la base de données relationnelle et le *tagging* puissent se compléter. Alors que le *tagging* est idéal pour établir une vue d'ensemble d'un corpus, la recherche lexicale tend à éloigner le chercheur du contexte textuel. La base de données relationnelles reste utile pour décomposer des portions de textes selon des thématiques préalablement choisies ainsi que pour établir des listes de noms en évitant l'écueil que peut poser une sur-automatisation de la recherche. Le risque est d'amener le chercheur à travailler

---

[51] http://shamela.ws. Cette base de données est réalisée en Arabie saoudite.

Adrien de Jarmy
Doctorant Sorbonne Université, Orient & Méditerranée, UMR 8167

de manière « hors-sol », en ne faisant que survoler les sources et en attribuant un sens unique à chaque terme du lexique collecté automatiquement, sans en étudier la place dans le texte. Dans tous les cas, les compétences numériques facilitent aujourd'hui grandement le travail de l'historien, et la réalisation de bases de données encourage le chercheur à développer une réflexion épistémologique sur ses sources qui ne peut être que bienvenue.

Adrien de Jarmy
Doctorant Sorbonne Université, Orient & Méditerranée, UMR 8167

# Annexes

Annexes 1. Structure générale de la base de données relationnelle en SQL

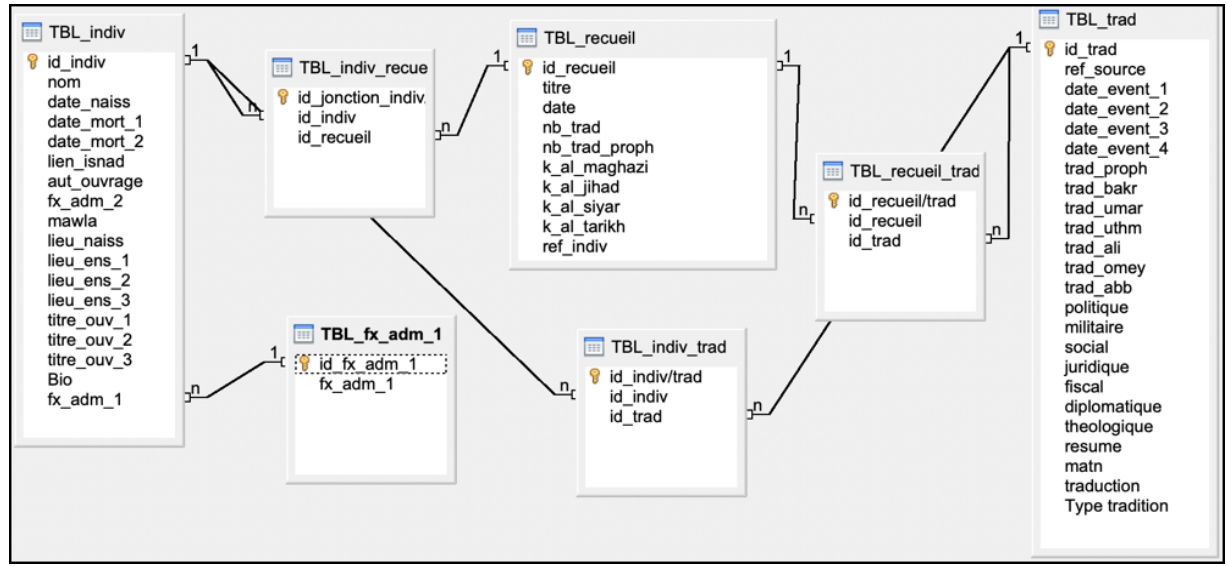

- TBL–indiv : table à visée prosopographique servant à la recension du nom et des informations personnelles de tous les individus rencontrés dans les listes des garants (*isnād*).
- TBL–recueil : table servant à la recension des différentes œuvres du corps.
- TBL–trad : table servant à la recension du contenu textuel des traditions (*matn*), sous la forme d'un texte pour résumer l'information et de catégories à cocher.
- TBL–fx–adm–1 : table permettant de déplier une liste des différentes fonctions administratives et politiques qu'un individu a pu occuper au cours de sa vie.
- TBL–indiv–recueil : table de liaison permettant de connecter les tables TBL–inv et TBL–recueil dans l'objectif d'effectuer des requêtes croisées entre ces tables.
- TBL–indiv–trad : table de liaison permettant de connecter les tables TBL–inv et TBL–trad.
- TBL–recueil–trad : table de liaison permettant de connecter les tables TBL–recueil et TBL–trad.

Adrien de Jarmy

Doctorant Sorbonne Université, Orient & Méditerranée, UMR 8167

Annexe 2. Présentation des liens entre les bases de données sous la forme d'un tableau chiffré

| id_indiv/trad | id_indiv | id_trad |
|---|---|---|
| 1 | 3 | 1 |
| 2 | 45 | 1 |
| 3 | 3 | 2 |
| 4 | 30 | 2 |
| 5 | 3 | 3 |
| 6 | 45 | 3 |
| 7 | 3 | 4 |
| 8 | 45 | 4 |
| 9 | 3 | 5 |
| 10 | 31 | 5 |
| 11 | 32 | 5 |

Annexe 3. Exemple de catégories booléennes

| | | | |
|---|---|---|---|
| trad_proph | ✔ | miracle | ☐ |
| trad_bakr | ☐ | diplomatique | ☐ |
| trad_umar | ☐ | theologique | ✔ |
| trad_uthm | ☐ | politique | ☐ |
| | | militaire | ☐ |

Ainsi la tradition listée numéro 1 dans la base de données (id–trad 1) a été transmise par deux individus (numéros 3 et 45), chaque numéro de la table id–indiv correspond à un nom précis. Les numéros de la colonne id–indiv/trad correspondent à la table de liaison TBL–indiv–trad qui relie ensemble les données des tables TBL–indiv et TBL–trad.

Adrien de Jarmy
Doctorant Sorbonne Université, Orient & Méditerranée, UMR 8167

Annexe 4. Exemple de catégorie à choix multiples : une liste de noms à sélectionner pour déterminer l'*isnād* d'une tradition.

Annexe 5. Résultat d'une requête dans la base de données (en anglais) afin de délimiter le pourcentage de traditions mentionnant le Prophète dans un échantillon de livres de *maġāzī*, à l'échelle de l'œuvre et des chapitres qui composent les œuvres.

| Collection and traditionist | Number of traditions in the whole text (*kitāb*) | Number of traditions mentioning Muḥammad in the text | Number of chapters (*bāb*) in the text | Number of chapters mentioning Muḥammad in the text | Percentages of traditions mentioning Muḥammad in the text | Percentages of chapters mentioning Muḥammad (at least once)) |
|---|---|---|---|---|---|---|
| *Muṣannaf* of ʿAbd al-Razzāq | 147 | 91 | 31 | 27 | 61,9% | 73% |
| *Muṣannaf* of Ibn Abī Šayba | 579 | 319 | 47 | 47 | 55,1% | 100% |
| *Ṣaḥīḥ* of al-Buḫārī | 488 | 402 | 90 | 87 | 88,3% | 96,6% |